# Silicon nanostructures embedded in superconductor shells

Nikolay T. Bagraev[1], Wolfgang Gehlhoff[2], Leonid E. Klyachkin[1], Anna M. Malyarenko[1],
Vladimir V. Romanov[3] and Serguey A. Rykov[3]

[1]*A.F. Ioffe Physico-Technical Institute, Russian Academy of Sciences, 194021 St. Petersburg, Russia*
[2]*Institut fuer Festkoerperphysik, Technische Universitaet Berlin, D-10623 Berlin, Germany*
[3]*St. Petersburg Polytechnical University, 195251 St. Petersburg, Russia*

We present the findings of the superconductivity in the silicon nanostructures prepared by short time diffusion of boron on the n-type Si (100) surface. These Si-based nanostructures represent the p-type ultra-narrow self-assembled silicon quantum wells confined by the δ - barriers heavily doped with boron. The resistivity, thermo-emf and magnetic susceptibility studies show that the high temperature superconductivity observed seems to result from the single-hole tunnelling into the negative-U boron centres at the silicon quantum well – δ-barrier interfaces.

Semiconductor silicon is well known to be the principal material for micro - and nanoelectronics. Specifically, the developments of the silicon planar technology are a basis of the metal-oxygen-silicon (MOS) structures and silicon-germanium (Si-Ge) heterojunctions that are successfully used as elements of modern processors [1]. Just the same goals of future high frequency processors especially to resolve the problem of quantum computing are proposed to need the application of the superconductor nanostructures that represent the Josephson junction series [2]. Therefore the manufacture of superconductor device structures in frameworks of the silicon planar technology seems to give rise to new generations in nanoelectronics. Furthermore, one of the best candidate on the role of the superconductor silicon nanostructure appears to be the δ - barriers heavily doped with boron that confine the high mobility silicon quantum wells (Si-QW) of the p-type located on the n-type Si (100) surface [3], because recently the heavily boron doping has been found to assist the superconductivity in diamond [4]. Here we present the first findings of the electrical resistivity, thermo-emf and magnetic susceptibility measurements that are actually evidence of the superconductor properties for the δ - barriers heavily doped with boron, which are revealed at high density of holes in the Si-QW. These silicon nanostructures embedded in superconductor shells are shown to be type II high temperature superconductors (HTS) with $T_c$=145 K and $H_{c2}$=0.22 T. Finally, the HTS silicon nanostructures are very promised to be applied as the sources and recorders of the THz irradiation by controlling the vortex transport phenomena.

The preparation of oxide overlayers on silicon monocrystalline surfaces is known to be favourable to the generation of the excess fluxes of self-interstitials and vacancies that exhibit the predominant crystallographic orientation along a <111> and <100> axis, respectively (Fig. 1a) [3,5,6]. In the initial stage of the oxidation, thin oxide overlayer produces excess self-interstitials that are able to create small microdefects, whereas oppositely directed fluxes of vacancies give rise to their annihilation (Figs. 1(a) and 1(b)). Since the points of outgoing self-interstitials and incoming vacancies appear to be defined by the positive and negative charge states of the reconstructed silicon dangling bond [6,7], the dimensions of small microdefects of the self-interstitials type near the Si (100) surface have to be restricted to 2 nm. Therefore, the distribution of the microdefects created at the initial stage of the oxidation seems to represent the fractal of the Sierpinski Gasket type with the built-in self-assembled Si-QW (Fig. 1(b)). Then, the fractal distribution has to be reproduced by increasing the time of the oxidation process, with the $P_b$ centres as the germs for the next generation of the microdefects (Fig. 1(c)) [7,8]. Although Si-QWs embedded in the fractal system of self-assembled microdefects are of interest to be used as a basis of optically and electrically active microcavities in optoelectronics and nanoelectronics, the presence of dangling bonds at the interfaces prevents such an application. Therefore, subsequent short-time diffusion of boron would be appropriate for the passivation of dangling bonds and other defects created during previous oxidation of the Si (100) surface thereby assisting the transformation of the arrays of microdefects in the neutral δ - barriers confining the ultra-narrow, 2nm, Si-QW (Fig. 1(d)).

We have prepared the p-type self-assembled Si-QWs with different density of holes ($10^9 \div 10^{12}$ cm$^{-2}$) on the Si (100) wafers of the n-type in frameworks of the conception discussed above and identified the properties of the two-dimensional high mobility gas of holes by the cyclotron resonance (CR), Hall-effect and infrared Fourier spectroscopy methods [3,5]. Besides, the secondary ion mass spectroscopy (SIMS) and scanning tunneling microscopy (STM) studies have shown that the δ - barriers, 3 nm, heavily doped with boron, $5 \cdot 10^{21}$ cm$^{-3}$, represent really alternating sequences of undoped tetrahedral microdefects and doped dots with dimensions restricted to 2 nm (Fig. 2, see also Fig. 1(d)). The analysis of the STM image enables to hazard a conjecture that this dimension of the microdefect of the self-interstitials type observed is consistent with the parameters expected from the tetrahedral model of the $Si_{60}$ cluster [9]. The value of the boron concentration determined by the SIMS method seems to indicate that each doped dot located between tetrahedral microdefects contains two impurity atoms of boron. Since the boron dopants form shallow acceptor centres in the silicon lattice, such high concentration has to cause a metallic-like conductivity. The boron centres packed up in dots have not been found, however, to result in an insulator-metal transition.

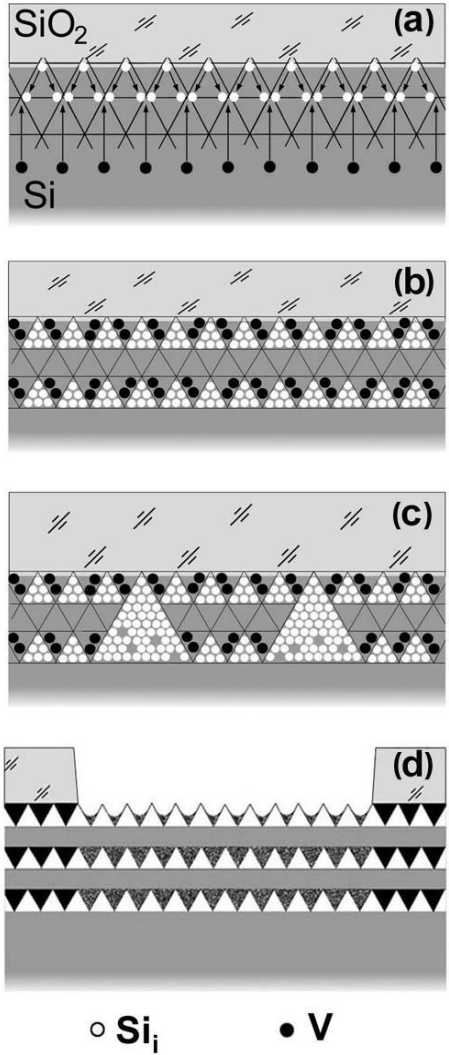

FIG. 1. A scheme of self-assembled silicon quantum wells confined by the δ - barriers heavily doped with boron on the Si (100) surface.
(a) The white and black balls label the self-interstitials and vacancies forming the excess fluxes oriented crystallographically along a <111> and <100> axis that are transformed to small microdefects.
(b) The silicon quantum wells between the layers of microdefects are produced by performing thin oxide overlayer.
(c) Besides, medium and thick oxide overlayers give rise to the reproduction of the fractal distribution of the microdefects of the self-self-interstitials type.
(d) In the process of short-time diffusion after making a mask and performing photolithography the atoms of boron replace the positions of vacancies thereby passivating the layers of microdefects and forming the neutral δ - barriers.

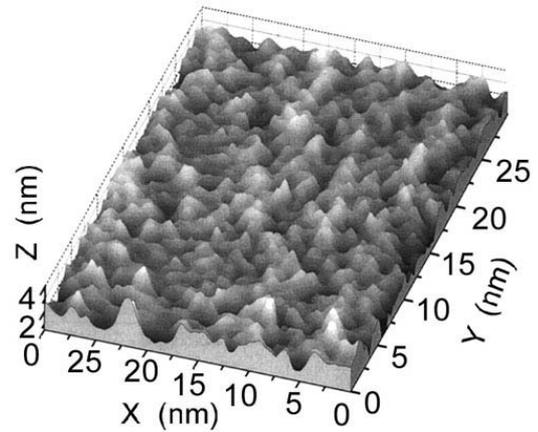

FIG. 2. Scanning tunneling microscopy image of the upper δ - barrier heavily doped with boron that confines the p-type silicon quantum well prepared on the n-type Si (100) surface. Light tetrahedrons result from the microdefects of the self-interstitials type. X∥[001], Y∥[010], Z∥[100].

The fall in the electrical activity of shallow boron acceptors appears to be attributable to the trigonal boron-related centres identified by the electron spin resonance method [3]. No ESR signals in the X-band are observed, if the Si-QW confined by the δ - barriers is cooled down in the external magnetic field ($B_{ext}$) weaker than 0.22 T, with the persistence of the amplitude and the resonance field of the trigonal ESR spectrum as function of the crystallographic orientation and the magnetic field value during cooling down process at $B_{ext} \geq 0.22$ T. Therefore the trigonal ESR spectrum observed seems to be evidence of the dynamic magnetic moment that is induced by the exchange interaction between the dipole boron centres, $B^+ - B^-$, which are formed by the negative-U reconstruction of the shallow boron acceptor pairs, $2B^0 \rightarrow B^+ + B^-$, along the <111> crystallographic axis. In common with the other solids that consist of small bipolarons [10], the δ - barriers containing the dipole boron centres have been found to be in an excitonic insulator regime at the density of holes in the Si-QW lower than $10^{11}$ cm$^{-2}$ [3,5]. The conductance of these silicon nanostructures appeared to be determined by the parameters of the 2D gas of holes in the Si-QW.

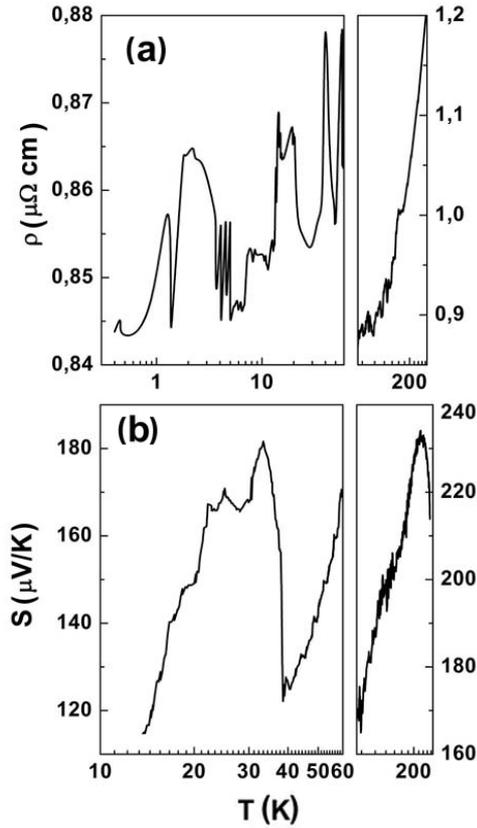

FIG. 3. Temperature dependences of electrical resistivity (a) and thermo-emf (Seebeck coefficient) (b) revealed by the B-doped silicon nanostructure under conditions of field-out cooling. The resistive and thermo-emf minima are used to define $\Delta$ and the values of the correlation gaps.

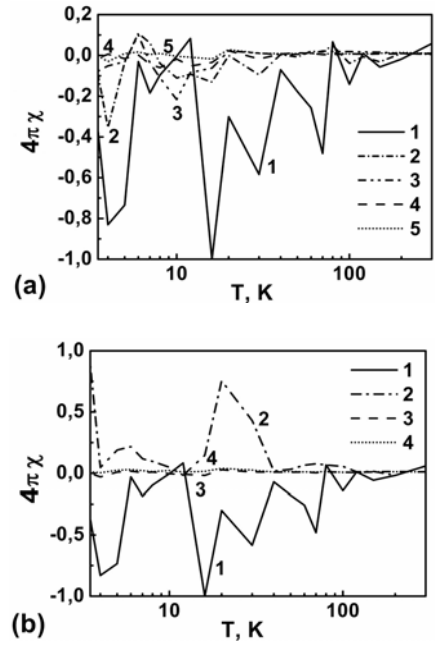

FIG. 4. Static magnetic susceptibility vs temperature that was revealed by the B-doped silicon nanostructure under conditions of field-in cooling.
a - diamagnetic response after field-out; 1 - 10 mT; 2 – 20 mT, 3 – 30 mT; 4 – 40 mT; 5 – 50 mT.
b - paramagnetic response after field-in (2, 4) and diamagnetic response after field-out (1, 3); 1,2 - 10 mT. 3,4 - 50 mT.

Here demonstrate that the high density of holes in the Si-QW ($\geq 10^{11}$ cm$^{-2}$) gives rise, however, to the superconductivity of the $\delta$ - barriers thereby forming very surprisingly silicon nanostructures inside the superconductor shells.

The current-voltage characteristics (CV) of the nanostructure with high density of holes in the Si-QW, $6 \cdot 10^{11}$ cm$^{-2}$, measured at different temperatures exhibited an ohmic character, whereas the temperature dependence of the resistivity, which is related to two-dimensional metal in the range 250-300 K, below 250 K reveals the behavior of inhomogeneous superconductor structure, with maxima and minima of peaks being in a good agreement with the thermo-emf features (Figs. 3(a) and 3(b)). The creation of each peak with decreasing temperature shows the logarithmic temperature dependence that appears to be caused by the Kondo-liked scattering of the single holes in SQW on the fluctuations due to inhomogeneous distribution of the negative-U boron dipoles in the $\delta$ - barriers. The Kondo-liked scattering seems to be the precursor of the optimal tunneling of single holes into the negative-U boron centers revealed by the positions of temperature minima of the resistivity and thermo-emf [11] (Figs 3(a) and 3(b)). This process is related to the conduction electron tunneling into the negative-U centers that is favourable to the increase of the superconducting transition temperature, $T_C$, in metal-silicon eutectic alloys [12,13]. The effect of single-hole tunneling is also possible to resolve some bottlenecks in the bipolaron mechanism of the high temperature superconductivity, which results from the lesser distance between the negative-U centres than the coherence length [14,15].

As was to be expected, the application of external magnetic field results in the shifts of the resistivity drops to lower temperatures, with the transformation to the logarithmic temperature increase in the range 0.4 - 100 K at $B_{ext} \geq 0.22$ T. The value of $\Delta=0.022$eV, $2\Delta=3.52kT_C$, $T_C=145$ K, as well as a series of correlation gaps that were derived from the measurements of the resistivity and thermo-emf appear to be revealed also in the temperature and magnetic field dependencies of the static magnetic susceptibility obtained by the Faraday method with the magnetic balance spectrometer MGD312FG (Figs. 4(a), 4(b) and 5). The value of temperatures corresponding to the drops of the diamagnetic response on cooling is of importance to coincide with the maxima of the resistivity and thermo-emf peaks thereby confirming the role of the charge correlations localized at the negative-U centres in the Kondo-liked scattering and the enhancement of $T_C$ (Fig. 4(a)). Just the same temperature dependence of the paramagnetic response observed after the field-in procedure exhibits the effect of the arrays of the Josephson transitions revealed by the STM image (Fig. 2) on the flux pinning processes in the superconductor δ - barriers heavily doped with boron (Fig. 4(b)). Besides, the preliminary studies of the specific heat in the temperature interval from 77 K to 300 K are found to demonstrate the corresponding jumps also at the temperatures near the maxima of the diamagnetic response.

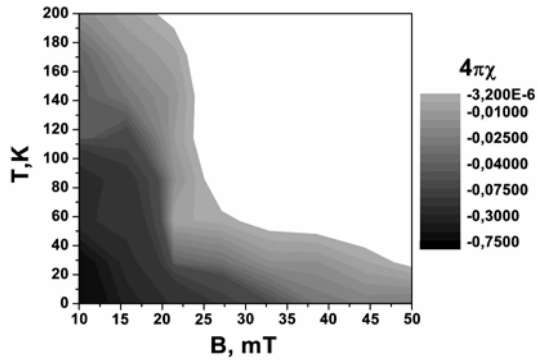

FIG. 5. Plots of static magnetic susceptibility vs temperature and magnetic field that was observed under conditions of field-in cooling in the B-doped silicon nanostructure. These data can be used visually to define $H_{c1}$, $H_{c1}$=215 Oe.

The plots of the magnetic susceptibility vs temperature and magnetic field result in the value of the coherence length, $\xi$=39 nm; where $\xi = (\Phi_0/2\pi H_{C2})^{1/2}$, $\Phi_0 = h/2e$; $H_{C2}$ is the second critical magnetic field ($H_{C2}$=0.22 T). This value of the coherence length appears to be in a good agreement with the values of Δ and the first critical magnetic field, $H_{C1}$=215 Oe (Fig. 5).

Therefore, apart from the bipolaron superconductivity, the results obtained have a bearing on the other versions of the high temperature superconductivity [16], specifically, based on the promising application of the sandwiches that consist of the alternating superconductor and insulator layers [17]. In the latter case, the sequences of the heavily doped and undoped silicon nanostructures are of advantage to achieve the high value of $T_c$, $T_c \approx \Theta \cdot \exp[-N(0) \cdot V]$, because of the high Debye temperature, Θ=625 K, that compensates for relatively low density of states, N(0).

Finally, the combinations of the δ - barriers heavily doped with boron and the p-type Si-QW prepared on the n-type Si (100) surface seem to be perspective for the physics of vortex as the sources and recorders of the THz irradiation. The electroluminescence spectra of the samples studied that have been obtained with the infrared Fourier spectrometer IFS-115 Brucker Physik AG reveal several modes of the THz irradiation in the interval from 0.12 THz to 5.3 THz. Thus, the best compromise on the presence of the $p^+$-n junction revealed by the CV characteristics and the control of the Rashba spin-orbit interaction by the gate-voltage applied to the superconductor – Si-QW sandwich opens up new possibilities for the transport of vortices, even though the external magnetic field is absent.